\documentclass[twocolumn]{tum-article}

\usepackage{siunitx}
\usepackage{cite}

\definecolor{cutomgreen}{rgb}{0.13, 0.55, 0.13}

\title{Intensity Pattern Types in Broadband Fourier Domain Mode-Locked (FDML) Lasers Operating Beyond the Ultra-Stable Regime}
\titlerunning{}
\authorrunning{Mark Schmidt et al.}

\author{Mark Schmidt\authormark{1,\Letter}\orcid{0000-0002-0934-5393}, Christin Grill\authormark{2}, Simon Lotz\authormark{2}, Tom Pfeiffer\authormark{2}\orcid{0000-0001-7791-1126}, Robert Huber\authormark{2}\orcid{0000-0001-9785-6574} and Christian Jirauschek\authormark{1}\orcid{0000-0003-0785-5530}}

\email{mark.schmidt@tum.de}

\affil[1]{TUM Department of Electrical and Computer Engineering, Technical University of Munich (TUM), Arcisstr. 21, 80333 Munich, Germany}
\affil[2]{Institute of Biomedical Optics, University of Lübeck, Peter-Monnik-Weg 4, 23562 Lübeck, Germany}

\date{}
\date{This is a post-peer-review, pre-copyedit version of an article published in Applied Physics B: Lasers and Optics (Appl. Phys. B \textbf{127}, 60, 2021). The final authenticated version is available online at: \url{https://doi.org/10.1007/s00340-021-07600-1}. }

\begin{document}

\maketitle

\begin{abstract}
	We report on the formation of various intensity pattern types in detuned Fourier domain mode-locked (FDML) lasers and identify the corresponding operating conditions. Such patterns are a result of the complex laser dynamics and serve as an ideal tool for the study of the underlying physical processes as well as for model verification. By numerical simulation we deduce that the formation of patterns is related to the spectral position of the instantaneous laser lineshape with respect to the transmission window of the swept bandpass filter. The spectral properties of the lineshape are determined by a long-term accumulation of phase-offsets, resulting in rapid high-amplitude intensity fluctuations in the time domain due to the narrow intra-cavity bandpass filter and the fast response time of the semiconductor optical amplifier gain medium.  Furthermore, we present the distribution of the duration of dips in the intensity trace by running the laser in the regime in which dominantly dips form, and give insight into their evolution over a large number of roundtrips.  
\end{abstract}

\section{Introduction}\label{sec_Introduction}
Fourier domain mode-locked (FDML) lasers produce rapidly wavelength-swept light with bandwidths of more than \SI{100}{\nano\meter} at tuning rates in the range of \SI{}{\mega\hertz}~\cite{huber2006buffered,wieser20105MHzOCT,klein2011megahertz,klein2013multi,kolb2016megahertz,kolb2018high}. This is achieved by synchronizing the roundtrip time of the optical field in a ring laser setup with the sweep rate of a tunable Fabry-P\'{e}rot (FP) filter, acting as the wavelength tuning element. The record sweep speeds and the excellent coherence properties have dramatically improved imaging and sensing applications, especially optical coherence tomography (OCT)~\cite{biedermann2009recent,reznicek2014megahertz}. The overall superior combination of tuning speed, imaging depth, sensitivity and axial resolution make FDML based OCT systems an interesting alternative to high-performance OCT systems, e.g.~\cite{hartl2001ultrahigh,bourquin2003ultrahigh,nishizawa2004real,leitgeb2004ultrahigh,wojtkowski2004ultrahigh,potsaid2010ultrahigh,nishizawa2018wavelength,wartak2020micro}.\\
Due to the large time-bandwidth product and the rapid time scales in optical systems in the order of \SI{}{\femto\second}, the analysis of the wavelength swept light in FDML lasers is a challenging task in general. 
Common experimental quantities of interest are the instantaneous frequency~\cite{butler2015single}, the instantaneous lineshape~\cite{butler2015single,biedermann2010direct,todor2011instantaneous}, and the intensity trace~\cite{kraetschmer2009ultrastable, slepneva2013dynamics,slepneva2019convective,pfeiffer2018ultra, lippok2019extended, jung2008characterization,marschall2010fourier}. Here, we focus on the intensity trace since it can be obtained by a relatively simple measurement with a photodiode and a real-time oscilloscope. We show that the intensity trace contains sufficient information to characterize the operation mode and the physical dynamics of the FDML laser, provided that sufficient analog bandwidth is available. Timing mismatches in FDML lasers can directly be observed in the intensity trace and are referred to as high-frequency fluctuations~\cite{pfeiffer2018ultra} since their time scale is much shorter than the sweep period.  
\\ It has been shown that the laser can operate without high-frequency fluctuations in the intensity trace in highly dispersion compensated and highly synchronized setups over a wavelength range of more than \SI{100}{\nano\meter}~\cite{pfeiffer2018ultra}. When timing delays, caused for example by either a residual dispersion in the fiber cavity or a detuning from the ideal sweep rate, exceed a certain amount, the intensity trace suffers from high-frequency fluctuations~\cite{pfeiffer2018ultra,schmidt2020self} which have a negative impact on the imaging quality in OCT applications. The high-frequency fluctuations can be classified as various types, such as irregular fluctuations, referred to as a modulational instability~\cite{slepneva2013dynamics,pimenov2017dispersive} and the Eckhaus instability~\cite{li2017eckhaus}, periodic Turing-type formations~\cite{slepneva2013dynamics,pimenov2017dispersive} and so-called holes~\cite{slepneva2019convective,pfeiffer2018ultra}.\\
Typically, the intensity trace of FDML lasers is recorded with measurement bandwidths of less than \SI{10}{\giga\hertz}~\cite{kraetschmer2009ultrastable, lippok2019extended, jung2008characterization,marschall2010fourier,jirauschek2009theoretical,wan2017time}, allowing to visualize only the low-frequency part. In this work, we characterize the intensity trace of an FDML laser at the full sweep speed using a real-time oscilloscope with a total analog bandwidth of several \SI{10}{\giga\hertz}, and report on the formation of characteristic intensity patterns under different operating conditions. The relatively large measurement bandwidth is essential to fully record the high-frequency fluctuations whose temporal extensions are limited by the bandwidth of the swept bandpass filter. Our results are further validated by numerical simulations which additionally make it possible to extract the spectral features of the operating modes. Interestingly, in many cases patterns are formed instead of irregular fluctuations even under the influence of strong detunings. The advantage of the method are its simplicity and the fact that variations of the optical phase are visible in the intensity trace as amplitude fluctuations.\\
The paper is organized as follows:
First, we describe the experimental setup and the underlying simulation model. Then we present the results and a comparison of experiment and simulation. In order to discuss specific properties of the high-frequency fluctuations, we demonstrate the distribution of the duration of so-called holes, i.e. dips in the intensity trace, and analyze the long-term evolution of the hole-type intensity patterns.        

\section{Intensity pattern types in non-synchronized FDML lasers} \label{sec_IntPatTypes}

The intensity trace of an FDML laser contains rich information about the interplay of the laser components due to the strong phase-amplitude coupling in the cavity. The extremely narrow bandpass filter introduces large losses on the optical field when the instantaneous wavelength of the optical field is offset from the central position in the spectral filter transmission window over many roundtrips. This occurs in the case of strong synchronization mismatches between the roundtrip time of the optical field and the sweep rate of the bandpass filter, caused by e.g. the fiber dispersion or a detuning from the ideal sweep rate. Typical values of the full width at half maximum (FWHM) bandwidth of the sweep filter are in the order of \SI{100}{\pico\meter} (several tens of \SI{}{\giga\hertz}). At a wavelength of \SI{1300}{\nano\meter} (\SI{231}{\tera\hertz}) and a filter bandwidth with a FWHM of \SI{165}{\pico\meter} (\SI{30}{\giga\hertz}), an offset of \SI{0.02}{\nano\meter} (\SI{15}{\giga\hertz}) from the peak transmission already causes a power loss of \SI{50}{\percent} when the reflected power is absorbed by an isolator. Such losses are compensated by the semiconductor optical amplifier (SOA) gain medium with a fast response time in the order of several tens of \SI{}{\pico\second}~\cite{schmidt2020self}. This interplay of frequency-shift, gain and loss over a long time scale, i.e. when this process is iterated over many roundtrips, introduces high-frequency fluctuations in the intensity trace with a large amplitude. In the following, we first  discuss the experimental laser setup as well as the simulation model, and present different types of high-frequency fluctuations by systematically detuning an FDML laser from the sweep rate at which the laser operates in the ultra-stable regime~\cite{kraetschmer2009ultrastable,pfeiffer2018ultra}.      

\subsection{Experimental setup}\label{subsec_ExpSetup}
\begin{figure*}[t]
	\resizebox{\textwidth}{!}{
		\centering
		\includegraphics{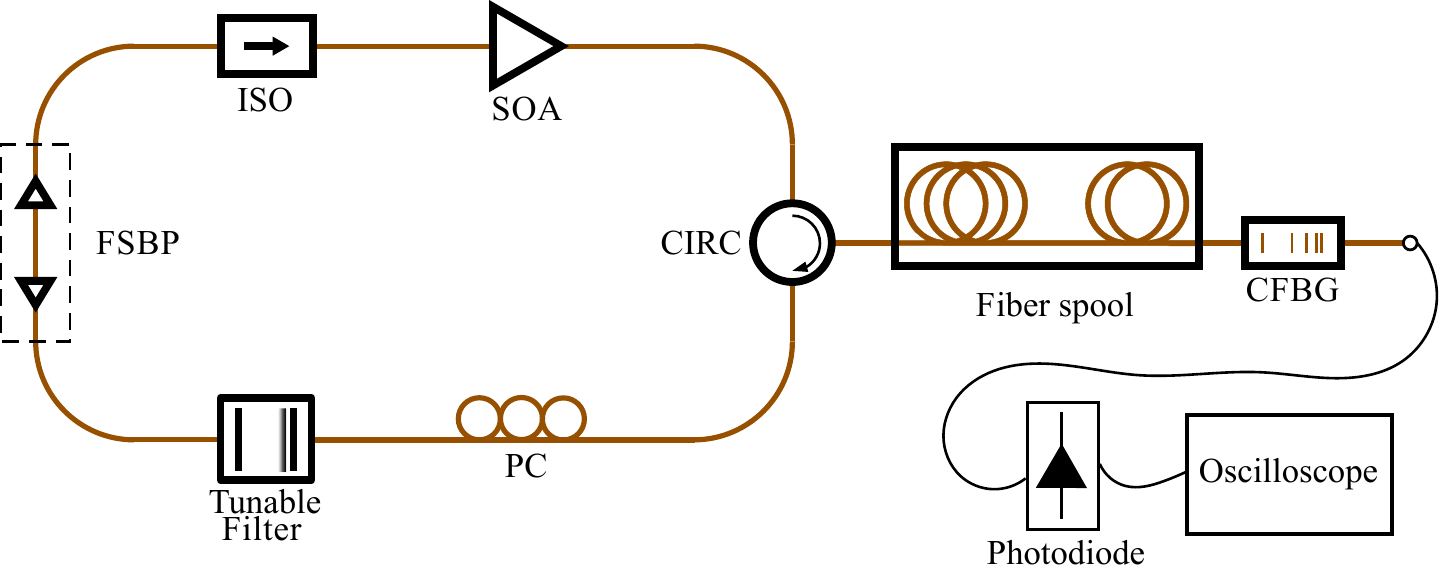}
	}
	\caption{Schematic illustration of the experimental setup.}
	\label{fig_Setup_Simon_Lotz}
\end{figure*}
We performed measurements with two similar FDML laser setups which differ in the output power, the bandwidth of the tunable bandpass filter and the mechanism to tune the sweep frequency. The first setup is described in detail in~\cite{pfeiffer2018ultra}. The second setup is illustrated in Fig.~\ref{fig_Setup_Simon_Lotz}. 
Here the sweep frequency of the tunable bandpass filter is kept constant, except when the laser is detuned, and the cavity length is tuned by modifying the length of a free space beam path (FSBP) in order to regulate the laser in the ultra-stable regime. The other components are a single polarization SOA gain medium (Thorlabs BOA1130S), a circulator (CIRC), a fiber spool consisting of a mix of SMF-28, HI1060 and LEAF fibers, and a temperature fine-tuned chirped fiber Bragg grating (cFBG) serves as a dispersion compensating element as well as the laser output (Teraxion, custom made). Further components are a polarization controller (PC) to adjust the polarization state of the light to the maximum gain axis of the SOA, a home-made tunable bandpass filter and an isolator (ISO) to ensure unidirectional lasing. The intensity trace was recorded with a \SI{33}{\giga\hertz} photodiode (Discovery semiconductors Inc.  DSC20H) whose output was digitized with a \SI{63}{\giga\hertz}, \SI{160}{\giga\siemens} real-time oscilloscope (Keysight DSOZ634A). The intensity traces referring to setup 1 are recorded with a \SI{50}{\giga\hertz} photodiode.\\
Unless otherwise mentioned, the data presented in the manuscript refers to the first setup. The data of Section \ref{sec_histogram} was generated with the second setup. No qualitative differences have been observed between the two setups, and thus the results presented here do not significantly depend on details of the underlying laser setup with respect to the parameters changed in this work. Typical values for the average power at the laser output are in the first setup \SI{\sim 30}{\milli\watt} and in the second setup \SI{\sim 7}{\milli\watt}. The bandwidth of the tunable bandpass filter is in the first setup \SI{165}{\pico\meter} and in the second setup \SI{290}{\pico\meter}, i.e. it differs by roughly a factor of \SI{2}{}.
\
\subsection{Simulation model}\label{subsec_SimModel}
In order to reproduce the experimental results, we performed simulations with the model presented in~\cite{schmidt2020self}. This model contains the fundamental interplay between the swept bandpass filter, a delay element such as a dispersive optical fiber or a detuning element and a single polarization SOA gain medium. The role of the simulation model in this work is twofold. First, by reproducing the measured intensity patterns, we can verify that the dynamics of the laser is sufficiently described by the three above mentioned processes, and we can identify the spectral features of the operating modes. Second, the simulation bandwidth is chosen to be \SI{3.45}{\tera\hertz} in order to show that no hidden intensity fluctuations exist which could not be captured with the limited measurement bandwidth of \SI{33}{\giga\hertz} and \SI{50}{\giga\hertz}, respectively.
\\
We added the effects of carrier heating (CH) and spectral hole burning (SHB) in the SOA gain medium to the model in~\cite{schmidt2020self}. A quasi-static approach is used, since these processes have time constants in the order of \SI{100}{\femto\second} which is much smaller than the temporal extension of the high-frequency fluctuations in FDML lasers. Within this approach the update equation for a lumped element SOA is given by 
\begin{align}\label{Eq_SOA_in_out}
u_\mathrm{out}(\tau) = \,\, & u_\mathrm{in}(\tau) \exp\left[ 0.5 h_\mathrm{tot}(\tau) \right] \nonumber \\
& \cdot   \exp\left\{-\mathrm{i} 0.5\left[ \alpha_\mathrm{N} h_\mathrm{N}(\tau) +\alpha_\mathrm{CH} h_\mathrm{CH}(\tau) \right] \right\},
\end{align} 
where $u_\mathrm{in,out}$ are the slowly varying complex electric field envelopes in the swept filter reference frame~\cite{jirauschek2009theoretical} at the spatial input and output of the SOA, respectively. The retarded time is denoted by $\tau$ and $\alpha_\mathrm{N,CH}$ are the linewidth enhancement factors due to band-filling and carrier heating. The total gain is given by $h_\mathrm{tot} = h_\mathrm{N} + h_\mathrm{CH} + h_\mathrm{SHB}$ with the contributions due to band-filling $(h_\mathrm{N})$, CH $(h_\mathrm{CH})$ and SHB $(h_\mathrm{SHB})$. The linewidth enhancement factor due to SHB $\alpha_\mathrm{SHB}$ is set to zero as in~\cite{mecozzi1997saturationFWM, wang2007temporal, zilkie2008time}. A description of the related differential equations of the three parts of the total gain is given in~\cite{cassioli2000time}. By setting $\partial h_{\mathrm{CH}}(\tau) / \partial \tau =  \partial h_{\mathrm{SHB}}(\tau) / \partial \tau \approx 0$, one can solve the resulting implicit equation with the principal branch of the Lambert W function $W_0$~\cite{corless1996lambertw} by defining a nonlinear gain $h_\mathrm{NL} = h_\mathrm{CH} + h_\mathrm{SHB}$, and one obtains
\begin{align}
h_{\mathrm{NL}} (\tau) = \,\, & \epsilon P_{\mathrm{in}}(\tau) - \tau_\mathrm{SHB} \left[ \partial h_{\mathrm{N}}(\tau) / \partial \tau \right] 
\nonumber \\
& - W_0\Big( \Big. \epsilon P_{\mathrm{in}}(\tau) \exp\left\{ \right. \epsilon P_{\mathrm{in}}(\tau) + h_{\mathrm{N}}(\tau) 
\nonumber \\
& - \tau_\mathrm{SHB} \left[ \partial h_{\mathrm{N}}(\tau) / \partial \tau \right] \left. \right\} \Big. \Big).
\label{Eq_Solution_NL}
\end{align}  
In Eq.~(\ref{Eq_Solution_NL}),  $\epsilon = \epsilon_{\mathrm{CH}} + \epsilon_{\mathrm{SHB}}$ is the combined gain compression factor where $\epsilon_{\mathrm{CH}}$ and $\epsilon_{\mathrm{SHB}}$ are the gain compression factors due to CH as well as SHB, and $P_{\mathrm{in}} = |u_\mathrm{in}|^2$ is the optical power at the spatial input of the SOA.
After computing the nonlinear gain $h_\mathrm{NL}$ with Eq.~(\ref{Eq_Solution_NL}), the individual components can then be found by 
\begin{align}
h_{\mathrm{SHB}} (\tau) &= -\epsilon_{\mathrm{SHB}} P_{\mathrm{in}}(\tau) 
\nonumber \\
& \quad \cdot \left\{ \exp \left[ h_{\mathrm{N}}(\tau) + h_{\mathrm{NL}}(\tau)\right] - 1 \right\} 
\nonumber \\
& \quad - \tau_\mathrm{SHB} \left[ \partial h_{\mathrm{N}}(\tau) / \partial \tau \right] 
\label{Eq_Solution_SHB},
\\ 
h_{\mathrm{CH}} (\tau) &= -\epsilon_{\mathrm{CH}} P_{\mathrm{in}}(\tau) 
\nonumber \\
& \quad \cdot \left\{ \exp \left[ h_{\mathrm{N}}(\tau) + h_{\mathrm{NL}}(\tau)\right] - 1 \right\}.
\label{Eq_Solution_CH}
\end{align}
The SOA parameters are taken from~\cite{cassioli2000time} and are $\tau_c = \SI{70}{\pico\second}$, $\alpha_\mathrm{N} = \SI{1.55}{}$, $\alpha_\mathrm{CH} = \SI{0.94}{}$, $\epsilon_\mathrm{CH} = \SI{1.95}{\per\watt}$, $\epsilon_\mathrm{SHB} = \SI{1.17}{\per\watt}$, and $\tau_\mathrm{SHB} = \SI{120}{\femto\second}$. The other parameters which are different from~\cite{schmidt2020self} are the dispersion coefficients of the optical fiber and the cFBG which are $\beta_{2,3,4},\hat{\beta}_{2,3,4} = 0 $ for setup 1 and 2. For setup 2 the power loss factor in the fiber spool $\kappa_f$ is $\SI{0.33}{}$, the reflectivity of the cFBG $R$ is $\SI{0.4}{}$, the center wavelength $\omega_c$ is $2\pi \cdot\SI{230.8}{\tera\hertz}$ $ (\SI{1300}{\nano\meter}) $, the sweep bandwidth $D_\omega$ is $2\pi \cdot \SI{17.8}{\tera\hertz}$ $(\SI{100}{\nano\meter})$, the FP filter bandwidth $\Delta_\omega$ is $2\pi\cdot \SI{50.7}{\giga\hertz}$ $(\SI{0.290}{\nano\meter})$, the FP filter transmission $T_\mathrm{max}$ is $\SI{0.075}{}$ including the losses of the FSBP as well as the PC, and the total cavity losses $L$ are $\SI{20}{\decibel}$.

\subsection{Detuning from the ultra-stable regime} \label{sec_Detuning_Exp}
\begin{figure*}[t]
	\resizebox{\textwidth}{!}{
		\centering
		\includegraphics{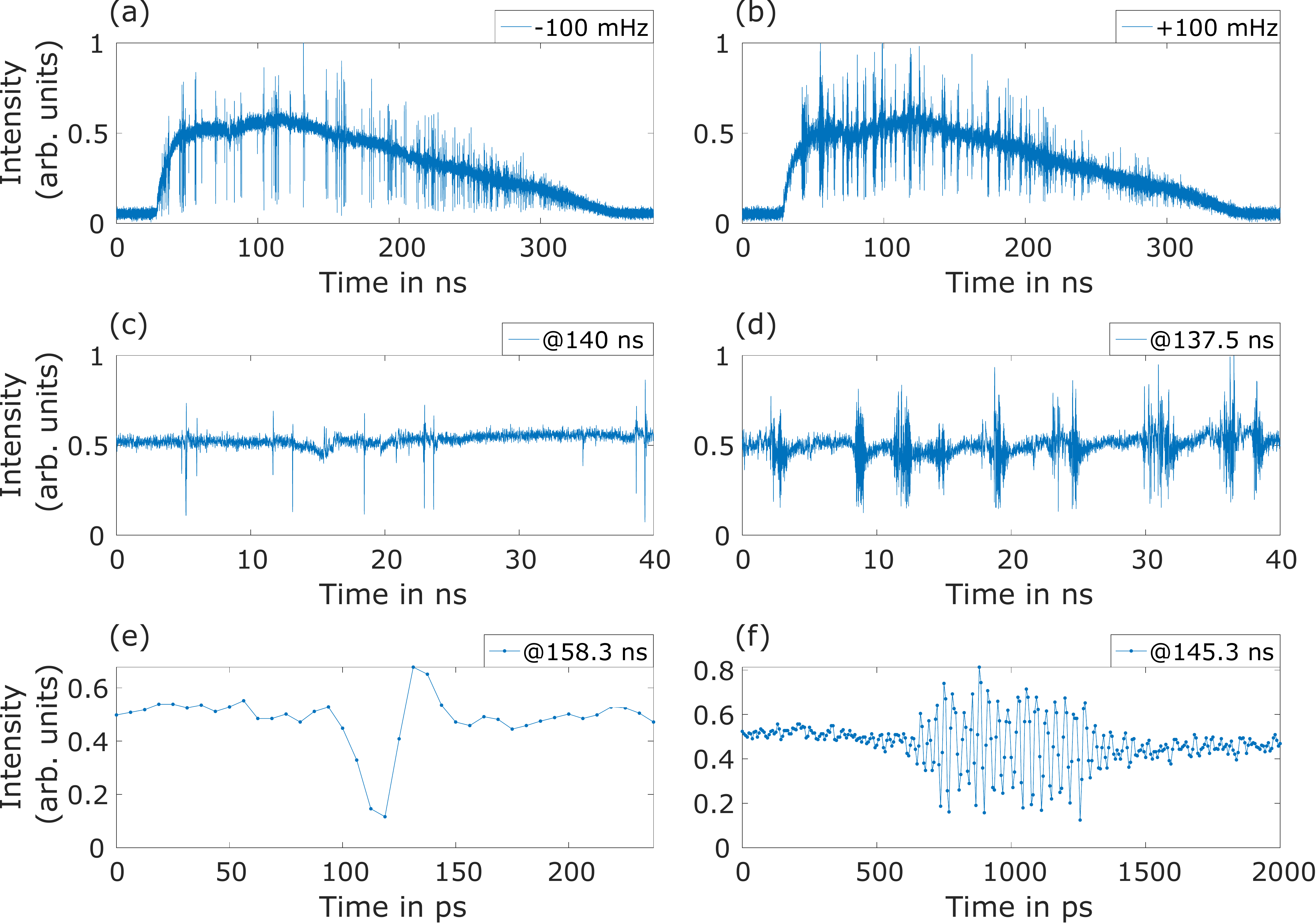}
	}
	\caption{(a) Measured intensity trace of a backward sweep. The laser is detuned by \SI{-100}{\milli\hertz} from the ultra-stable regime. (b) The same case as in (a) but with a positive sign in the detuning. (c) Zoom into an arbitrary position in the sweep which demonstrates that the intensity trace contains mainly hole-type patterns when the roundtrip time of the optical field is smaller than the sweep rate. (d) The same as in (c) but for a reversed sign in the detuning, demonstrating that fringes form instead of holes. (e) Single hole and (f) single fringe.}
	\label{fig_100mHz}
\end{figure*}
\begin{figure*}[t]
	\resizebox{\textwidth}{!}{
		\centering
		\includegraphics{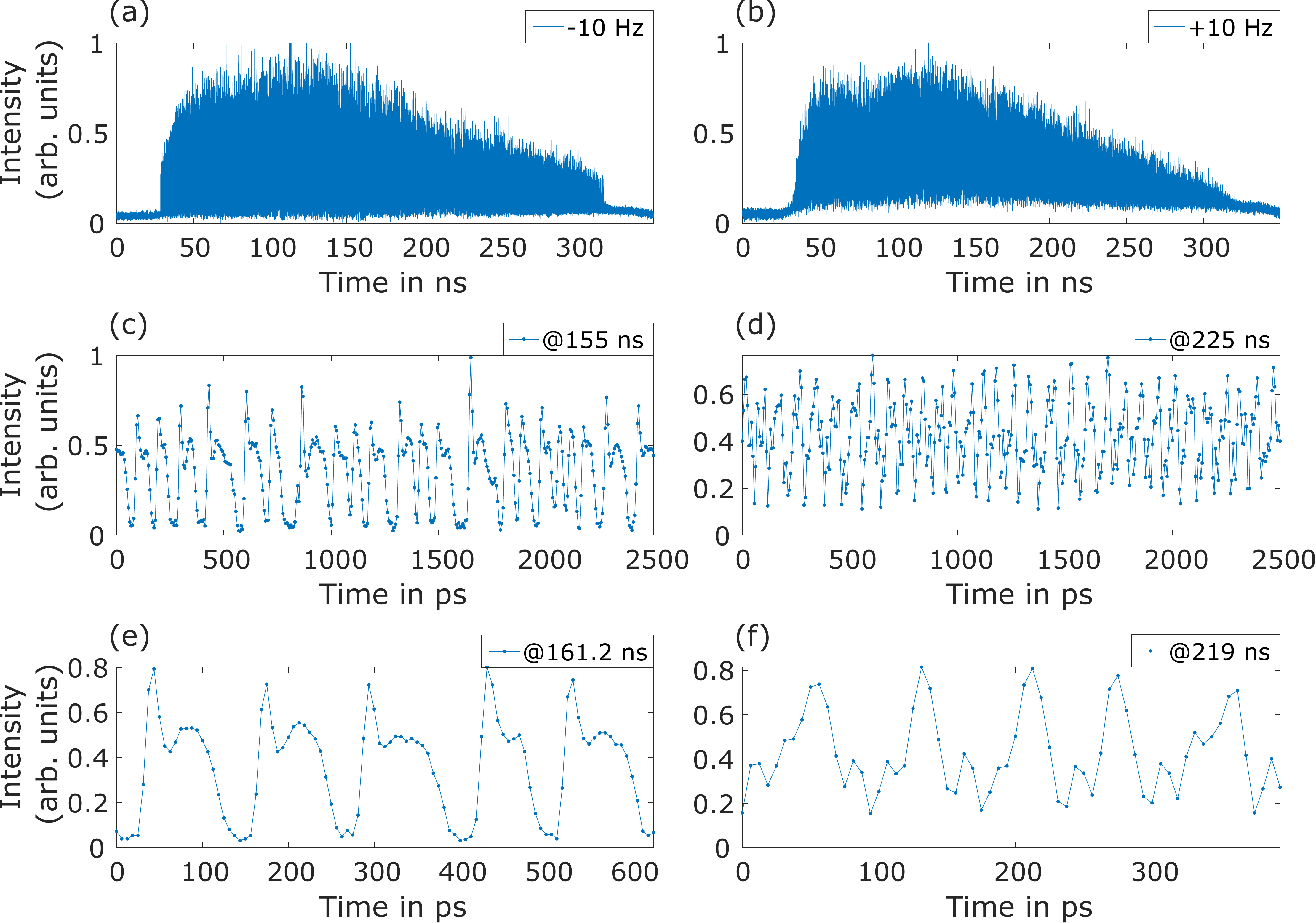}
	}
	\caption{Measured intensity trace of a backward sweep. The laser is detuned by (a) \SI{-10}{\hertz} from the ultra-stable regime and (b) by \SI[retain-explicit-plus]{+10}{\hertz}. (c) Zoom into an arbitrary position to demonstrate the existence of quasi-periodic patterns instead of irregular fluctuations. (d) Similar to (c) but for reversed sign in the detuning, resulting in different shape. (e) and (f) Zoom into five patterns with similar shape.}
	\label{fig_10Hz}
\end{figure*}
In Fig.~\ref{fig_100mHz} the experimental intensity traces of a laser detuned by \SI{+-100}{\milli\hertz} from the ultra-stable regime are shown at different time scales: The full sweep in Figs.~\ref{fig_100mHz} (a) and (b), an arbitrarily selected \SI{40}{\nano\second} window in Figs.~\ref{fig_100mHz} (c) and (d) and individual patterns in Figs.~\ref{fig_100mHz} (e) and (f), respectively. The sinusoidal filter sweep rate in both cases is \SI{411}{\kilo\hertz} and the gain medium is operated with a \SI{12.5}{\percent} duty cycle~\cite{pfeiffer2018ultra}, such that a single sweep lasts \SI{\approx  ~300}{\nano\second} where the sweep filter drive function is nearly linear. The sweep in this time window has a bandwidth of \SI{117}{\nano\meter} and is centered at a wavelength of \SI{1292}{\nano\meter}. Interestingly, the density of high-frequency fluctuations is similar in both cases in Figs.~\ref{fig_100mHz} (a) and (b), yet the shape of the emerging patterns is different. We collected a large number of measurements by systematically detuning the laser from the sweep rate in the ultra-stable regime and conclude that
\begin{itemize}
	\item moderate negative detuning of a backward sweep is dominated by holes, see Fig.~\ref{fig_100mHz} (a), (c) and (e), whereas
	\item moderate positive detuning of a backward sweep is dominated by localized fringes, see Fig.~\ref{fig_100mHz} (b), (d) and (f).
	\item In the case of a forward sweep the situation is reversed, i.e. holes occur at \SI[retain-explicit-plus]{+100}{\milli\hertz} and fringes at \SI{-100}{\milli\hertz}, respectively. 
	\item At strong detunings, short quasi-periodic patterns can occur at certain positions in the sweep with different shape in the backward and forward sweep, as discussed below. 
\end{itemize}
These patterns are a result of the long-term interplay of the swept bandpass filter, the SOA gain medium and the frequency shift per roundtrip caused by the detuning, as will be shown in Section~\ref{sec_CompToSimModel}. The qualitative shape of the patterns does not depend on the temporal position in the sweep.\\
Figure~\ref{fig_10Hz} presents a comparison to the same setup as above, however for strong detuning of \SI{+-10}{\hertz}. As can be seen in Figs.~\ref{fig_10Hz} (c) and (d), a pulsed output is generated rather than irregular fluctuations, which might be unexpected when looking at the full sweeps in Figs.~\ref{fig_10Hz} (a) and (b). Yet, only short quasi-periodic groupings of pattern happen, and in some parts of the intensity trace irregular fluctuations can occur. Therefore, a solitary solution does not exist in this particular setup. From a fundamental point of view, it is interesting if a periodic pulsed solution in FDML lasers can exist at all and if such a solution constitutes a new type of dissipative soliton~\cite{grelu2012dissipative}. 

\section{Comparison to the simulation model} \label{sec_CompToSimModel}
\begin{figure*}[t]
	\resizebox{\textwidth}{!}{
		\centering
		\includegraphics{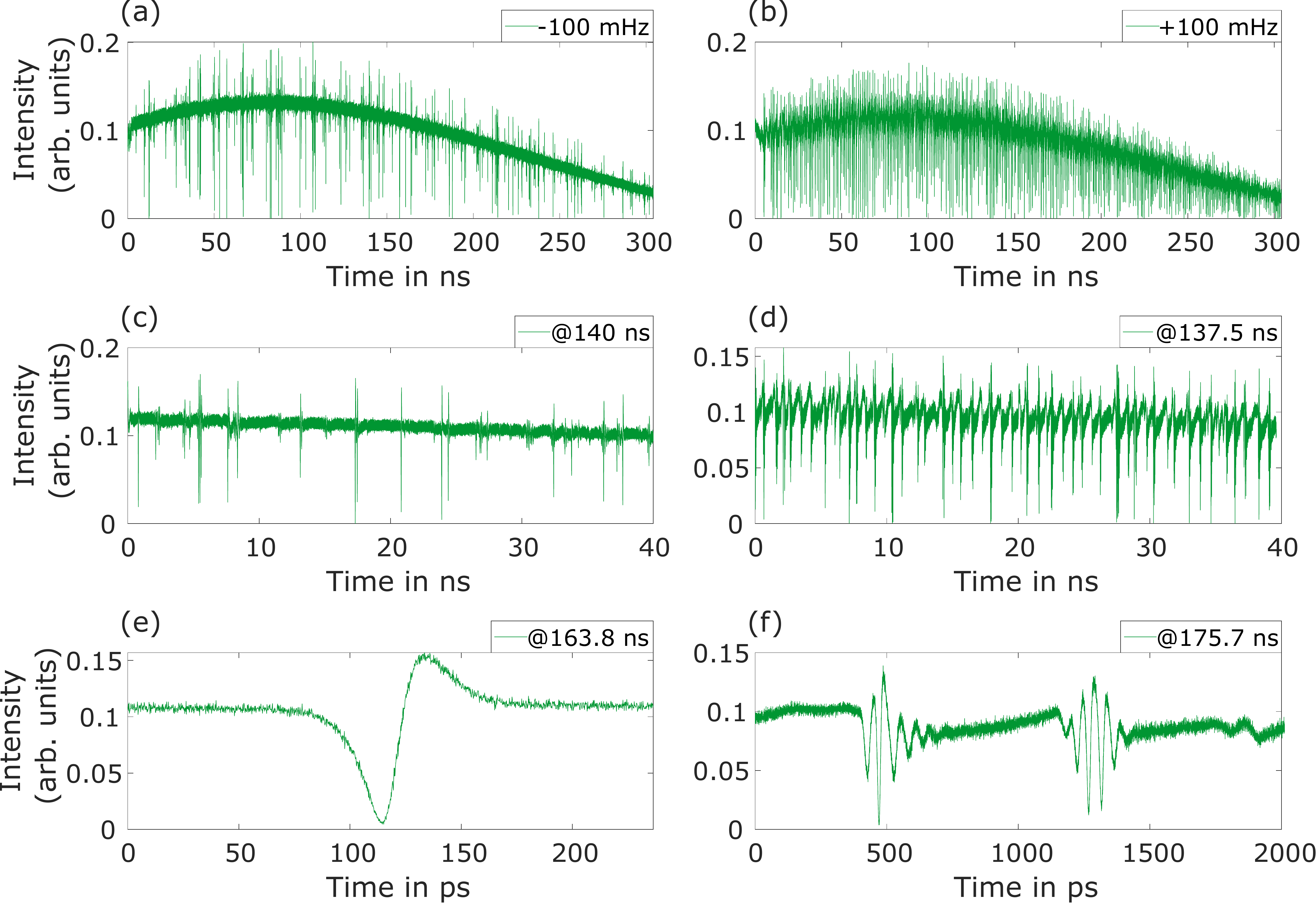}
	}
	\caption{(a) Simulated intensity trace of a laser which is detuned by \SI{-100}{\milli\hertz} as in Fig.~\ref{fig_100mHz}. (b) The same case as in (a) but with a different sign in the detuning. (c) Zoom into an arbitrary position in the sweep. (d) The same as in (c) but for reversed sign in the detuning, resulting in fringes instead of holes. (e) Extraction of a single hole and (f) of single fringes.}
	\label{fig_100mHz_simulatiuon}
\end{figure*}
\begin{figure*}[t]
	\resizebox{\textwidth}{!}{
		\centering
		\includegraphics{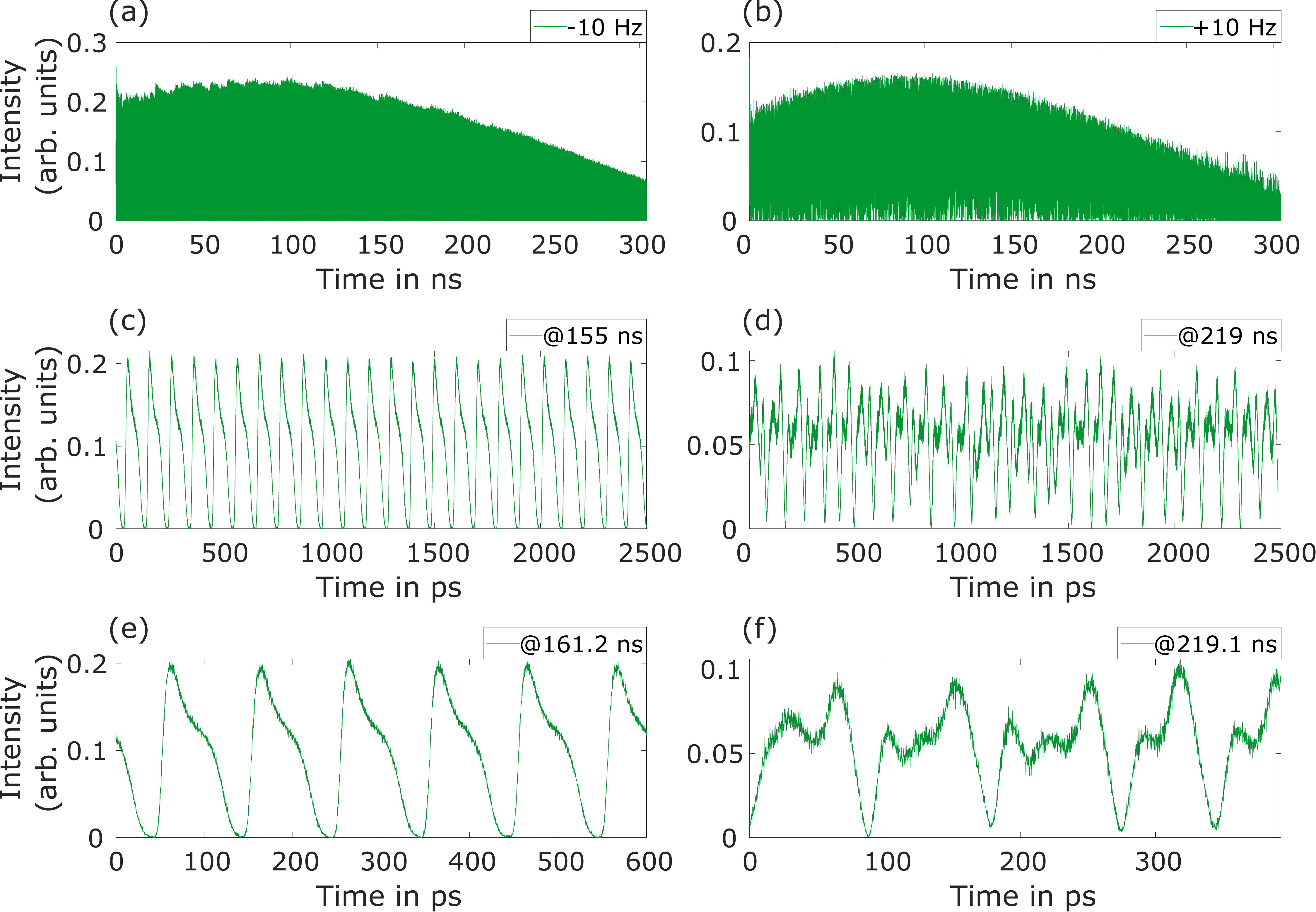}
	}
	\caption{Simulated intensity trace of a laser which is detuned by (a) \SI{-10}{\hertz} and (b) by \SI[retain-explicit-plus]{+10}{\hertz} as in Fig.~\ref{fig_10Hz}. (c) Zoom into a similar position as in the experimental case to demonstrate the existence of short-term quasi-periodic patterns instead of irregular fluctuations. (d) Similar to (c) but for reversed sign in the detuning, resulting in different shape. (e) and (f) Zoom into five patterns with similar shape.}
	\label{fig_10Hz_simulation}
\end{figure*}
We performed numerical simulations of the first setup in order to reproduce Figs.~\ref{fig_100mHz} and~\ref{fig_10Hz}. Here we neglected the dispersion in the fiber spool causing a maximum residual group delay of less than \SI{200}{\femto\second}~\cite{pfeiffer2018ultra}, since this contribution is small compared to the delay introduced by a detuning of \SI{+- 100}{\milli\hertz} (\SI{+- 10}{\hertz}) per roundtrip  which is \SI{ \sim 590}{\femto\second} (\SI{\sim 5.9}{\pico\second}) in the case of a linear sweep. Yet, differences in the hole density due to the unknown residual dispersion as well as residual temperature fluctuations in the experiment are to be expected, especially in the case of low detuning.\\
As can be seen from Figs.~\ref{fig_100mHz_simulatiuon} and \ref{fig_10Hz_simulation}, the simulation model reproduces the qualitative behavior of the experimental data in Section~\ref{sec_Detuning_Exp} extremely well and is therefore an excellent tool to study the FDML laser dynamics. Yet, differences in the exact shape exist in the case of the local fringes or the quasi-periodic patterns while holes can be reproduced almost exactly, see also~\cite{schmidt2020self}. 
This observation shows that the accuracy of the individual spectral shifts introduced by the SOA, the bandpass filter or the detuning are a bottleneck in modeling FDML lasers, which has not yet been fully addressed in literature in this detail, to the best of our knowledge. \\
In Figs.~\ref{fig_100mHz_simulatiuon} (d) and (f), the local fringes are reduced in length, and thus also the density differs by roughly a factor of three. For negative detuning, the quasi-periodic patterns have a similar shape as in experiment (see Fig.~\ref{fig_10Hz} (e)), but a higher periodicity in Fig.~\ref{fig_10Hz_simulation} (e). We found that individual shapes can indeed be reproduced by modifying the SOA parameters. In the case of Fig.~\ref{fig_10Hz_simulation} (e) a higher $\alpha_\mathrm{N}$ of around \SI{3}{} qualitatively destabilizes the strong periodicity, yielding a good match of the detailed shape with the experimental result in Fig.~\ref{fig_10Hz} (e). The local fringes in Fig.~\ref{fig_100mHz_simulatiuon} (f) and the pattern in Fig.~\ref{fig_10Hz_simulation} (f) agree well with experiment for $\tau_\mathrm{c} = \SI{380}{\pico\second}$ and $\alpha_N = \SI{5.0}{}$, but at the cost of a worse agreement in terms of the hole duration statistics, as will be discussed in Section~\ref{sec_histogram}. Generally speaking, the assumption of a static carrier lifetime $\tau_\mathrm{c}$ and linewidth enhancement factor $\alpha_\mathrm{N}$ limits the accuracy of the model~\cite{schmidt2020self,wang2007temporal,zilkie2008time},
but as presented in this work it is still an ideal tool to predict and study the dynamical processes of FDML lasers from a qualitative point of view. As can be speculated from the previous discussion, an optimal set of time independent SOA parameters exists with the best match to the experimental intensity pattern in the sense of e.g. an Euclidean distance~\cite{liao2005clustering}. Yet, due to the nonstationary nature of the intensity patterns, the development of a suitable cost function is a challenging task and from a physical point of view this approach is questionable anyway. Therefore, such a procedure is not followed in this work.
\\   
A considerable dependence of the pattern shapes on the gain compression parameters $\alpha_\mathrm{CH}$, $\epsilon_{\mathrm{CH,SHB}}$ or $\tau_\mathrm{SHB}$ could not be observed. In particular, the contribution of the last term in Eq.~(\ref{Eq_Solution_NL}) and Eq.~(\ref{Eq_Solution_SHB}) involving $\tau_\mathrm{SHB}$ is two orders of magnitude smaller than $h_\mathrm{SHB}$ and $h_\mathrm{CH}$. The dominant component of the total gain is the band-filling component $h_\mathrm{N}$, and the CH and SHB components add a perturbation to $h_\mathrm{N}$. According to Eq.~(\ref{Eq_SOA_in_out}), the intensity of the incoming field is modified by the gain compression terms according to $\exp\left[ h_\mathrm{NL} (\tau)\right]$, and the instantaneous frequency according to $(1/4/\pi)\alpha_\mathrm{CH} \partial h_\mathrm{CH}(\tau) / \partial \tau $. Note that a change in the instantaneous frequency of the optical field affects the intensity in subsequent roundtrips because of the frequency dependent loss induced by the bandpass filter. This effect is best observed in the hole duration statistics discussed in Section~\ref{sec_histogram}. The ratio $\allowbreak \exp\left[ h_\mathrm{NL} (\tau)\right] \allowbreak / \exp\left[ h_\mathrm{tot} (\tau)\right]$, computed from the individual gain terms stored in the simulation, yields the change in the intensity due to the gain compression terms. According to our simulation results, this change is in the range of \SI{1}{}--\SI{7}{\percent} for various sweeps and detunings. As observed by a visual comparison of the patterns of simulations with and without the CH and SHB terms, the shape of the presented patterns is not changed considerably by this modification, and for example in the case of the hole-type patterns, this difference can hardly be captured on a visual basis. Therefore, we refer to the hole duration statistics in Section~\ref{sec_histogram} for a discussion of the long-term effect.  
\subsection{Spectral signatures of the optical field related to the formation of intensity patterns}\label{subsec_SpecSig}
In the subsequent discussion, we address the spectral dynamics related to the formation of different patterns, such as holes or fringes in Fig.~\ref{fig_100mHz} and Fig.~\ref{fig_100mHz_simulatiuon}, with respect to the sign of the detuning from the ultra-stable regime. 
\begin{figure*}[t]
	\resizebox{\textwidth}{!}{
		\centering
		\includegraphics{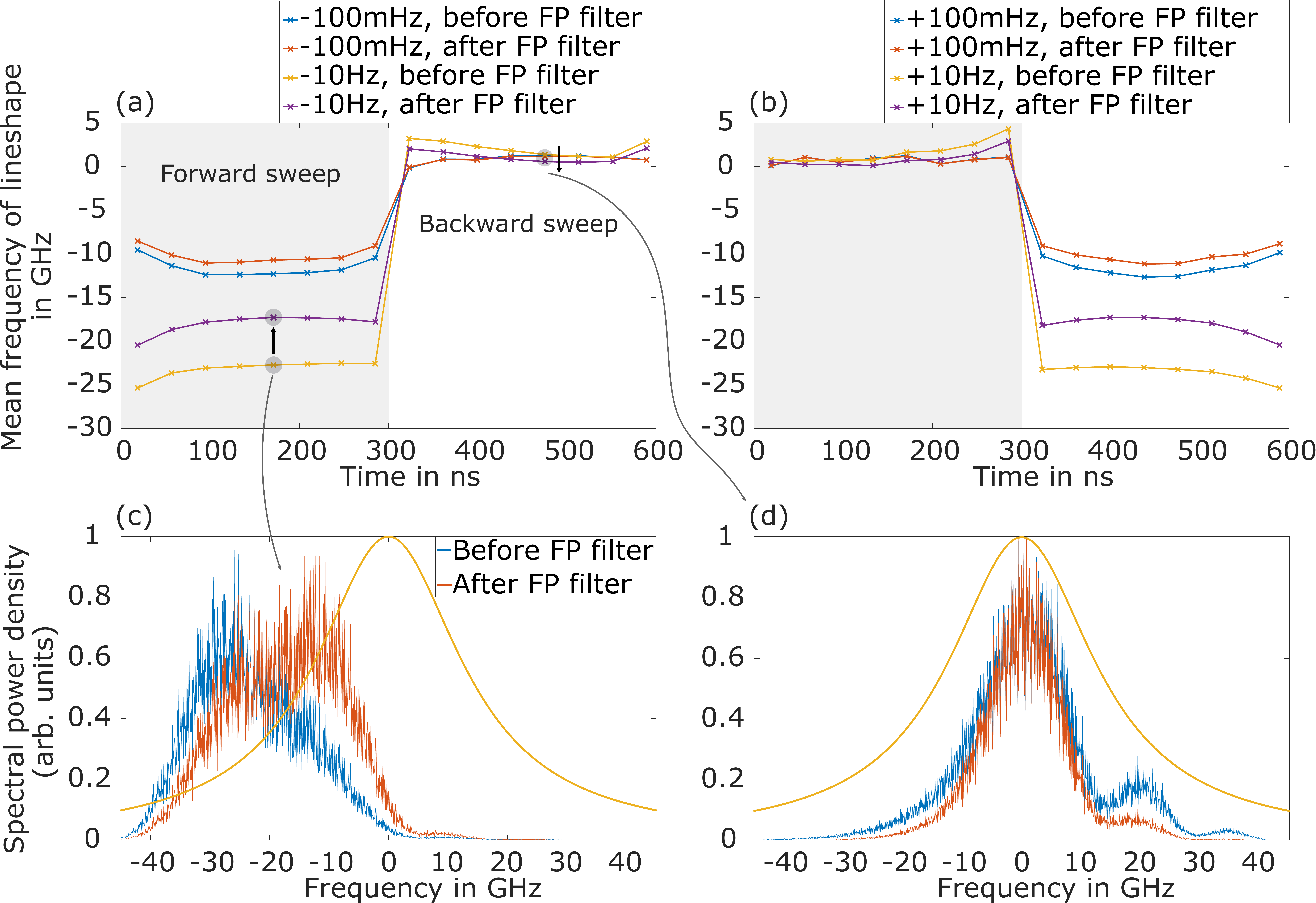}
	}
	\caption{(a) and (b) Mean frequency of the instantaneous lineshape for different strengths of (a) negative and (b) positive detunings. (c) and (d) Instantaneous lineshape at the spatial input and output position of the swept FP filter at the marked position (c) in the forward and (d) in the backward sweep in (a). The sweep filter transmission window is added in yellow for comparison in (c) and (d).}
	\label{fig_lineshape_comparison}
\end{figure*}
The formation of different patterns can be associated with different positions of the instantaneous lineshape in the spectral transmission window of the tunable bandpass filter. The instantaneous lineshape is defined and computed as in~\cite{schmidt2020self}, and this approach has also been used to reproduce measured lineshapes~\cite{todor2011instantaneous,todor2012balance}. The lineshape was averaged over 25 roundtrips to reduce spectral fluctuations. In Figs.~\ref{fig_lineshape_comparison} (a) and (b) the mean frequency of the instantaneous lineshape (MFL), or similarly the center of gravity, at the \SI{40000}{}\textsuperscript{th} roundtrip is plotted for both sweep directions and for different signs as well as strengths in detuning. This procedure is iterated at the spatial input and output of the swept FP filter. Note that the results refer to the field in the swept filter reference frame and the linewidth is not broadened by the sweep filter movement~\cite{todor2011instantaneous}. Clearly, each operating regime is associated with either a different position of the mean frequency with respect to the center of the swept bandpass filter, here at \SI{0}{\giga\hertz}, or a more pronounced spectral shift of the lineshape as in the case of the backward sweep in Fig.~\ref{fig_lineshape_comparison} (a) or forward sweep in Fig.~\ref{fig_lineshape_comparison} (b), respectively. To exemplarily illustrate the spectral shift of the lineshape by the FP filter, the transformation of the lineshape at the marked positions in Fig.~\ref{fig_lineshape_comparison} (a) for \SI{+-10}{\hertz} is shown in Fig.~\ref{fig_lineshape_comparison} (c) and (d).   
The simulation also confirms that the sign of the detuning simply interchanges the position of the mean frequency in the forward and backward sweep and therefore the type of pattern which dominates the intensity trace. \\
Furthermore, the position of the MFL with respect to the center of the transmission window of the swept bandpass filter is consistent with the sign of the frequency shift $\delta f$ caused by the detuning per roundtrip in the instantaneous frequency of the optical field. This frequency shift can approximately be computed by $\delta f = \tau_\mathrm{d}\, \partial \omega_s(\tau) / \partial \tau$~\cite{schmidt2020self}. Here $\tau_\mathrm{d}$ is the temporal delay of the optical field with respect to the center frequency of the swept bandpass filter, caused by a detuning from the filter sweep rate $f_0$ by the frequency $f_\mathrm{d}$. The center frequency of the swept bandpass filter is $\omega_s(\tau) + \omega_c $ with the center frequency of the sweep $\omega_c$ and $\omega_s(\tau + 1/f_0) = \omega_s(\tau)$. In our setup the relative sweep frequency $\omega_s$ is nearly linear in the time window where the gain medium is switched on, i.e. $\partial \omega_s / \partial \tau \allowbreak \approx \pm m D_\omega/T$ with $m=\allowbreak 1/\SI{12.5}{\percent}\allowbreak =8$ to account for the modulation of the gain medium. In the backward sweep $\partial \omega_s / \partial \tau > 0 $ and in the forward sweep $\partial \omega_s / \partial \tau < 0$ respectively. Note that the terminus forward or backward refers to the change in wavelength. Thus, the modified roundtrip time is $T + t_\mathrm{d}= 1/(f_0 + f_\mathrm{d})$ in the detuned case. The delay $\tau_\mathrm{d}$ is then given by $-f_\mathrm{d}T^2/(1 + f_\mathrm{d}T) \approx -f_\mathrm{d}T^2$~\cite{jirauschek2009theoretical}, since $ |f_\mathrm{d}T| \ll 1$ for typical sweep parameters. By combining the above equations we have
\begin{equation} \label{Eq_frequency_shift_detuning}
\delta f = -m \frac{f_\mathrm{d}}{f_0}D_\omega \mathop{\mathrm{sgn}}\left( \frac{\partial \omega_s}{\partial \tau}\right)
\end{equation} 
with the signum function $\mathop{\mathrm{sgn}}$.
We can now compute $\delta f $ for \SI{-+ 100}{\milli\hertz} which is \SI{-+41}{\mega\hertz} in the case of the forward sweep and \SI{+-41}{\mega\hertz} in the backward sweep case, respectively. The sign of $\delta f$ explains the position of the MFL in Fig.~\ref{fig_lineshape_comparison} (a) and (b) where in the case of $\delta f > 0$ the mean frequency is located near the center of the filter but is greater than zero most of the time. Interestingly, in~\cite{girard2011soa} an instable right hand side of the bandpass filter, which is in Figs.~\ref{fig_lineshape_comparison} (c) and (d) at $f > 0$, has been discussed in the context of SOA fiber ring lasers. Here, a combination of modulational instability and asymmetric four wave mixing between longitudinal cavity modes in the presence of the frequency dependent loss of the bandpass filter prohibits the instantaneous lineshape to stabilize at frequencies $f > 0$. Furthermore, the position of the MFL scales with the strength of the detuning as well as the magnitude of the spectral shift introduced by the FP filter as presented in Fig.~\ref{fig_lineshape_comparison}. Especially the formation of short quasi-periodic patterns is related to a large shift in the MFL compared to the case of localized fringes or hole patterns. The fact that the MFL is nearly constant in a single sweep agrees well with the observation in experiment that the type of pattern does not change over the full sweep.\\
In summary, we have shown that our simulation model reproduces the patterns in different operating modes on a qualitative basis and we discussed that the spectral position of the instantaneous lineshape in the sweep filter is strongly correlated with the type of pattern in the intensity trace. We also computed the spectral shift in the optical field introduced by the detuning and found consistent agreement between the sign of $\delta f$ and the position of the MFL with respect to the center of the filter transmission window. In addition, the symmetry between backward and forward sweep when the sign of the detuning is changed can be confirmed by an equivalent symmetry of the MFL.       

\section{Statistical evaluation of the hole duration} \label{sec_histogram}
\begin{figure}[t]
	\centering
	\resizebox{0.48\textwidth}{!}{
		\includegraphics{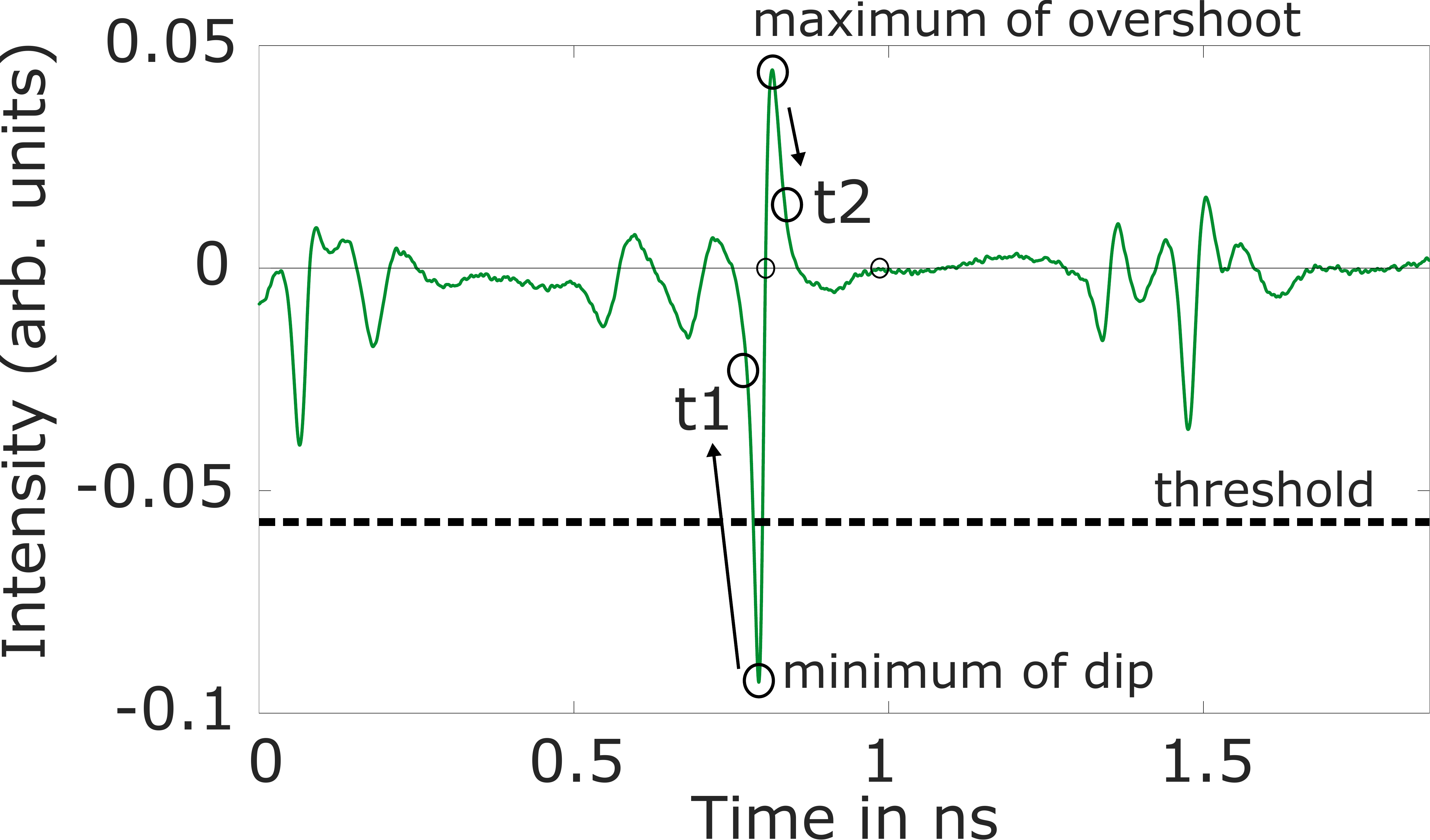}
	}
	\caption{Illustration of the algorithm to calculate the hole duration. The local mean is subtracted from the intensity trace. Therefore the zero-intensity line represents the reference level. The black dashed line is the threshold level, i.e. when the minimum of the dip deviates by more than \SI{50}{\percent} from the local mean.}
	\label{fig_hole_duration_illustration}
\end{figure}

The patterns discussed in Section~\ref{sec_IntPatTypes} are not stationary over a long time scale and the hole-type patterns in particular appear to propagate as well as modify their shape over successive roundtrips~\cite{pfeiffer2018ultra}. Furthermore, they are distorted by noise, such as fluctuations induced by the amplified spontaneous emission (ASE) noise of the SOA.\\ 
In this context, it is of practical as well as fundamental interest to extract the characteristic features of the individual patterns. Furthermore, this enables a systematic comparison to numerical simulations and contributes to a better visualization of their long time behavior. In the case of holes such as in Figs.~\ref{fig_100mHz} (a), (c) and (d), the temporal extension is an interesting fundamental parameter. Here, we describe a simple and robust algorithm to extract the hole duration from noisy data and present the distribution of the hole duration over a large number of roundtrips. \\
\textbf{Definition of the hole duration:} We define the hole duration $\Delta t_\mathrm{h}$ as $  t_2 - t_1$, where $t_1$ is the point in time when the intensity is \SI{75}{\percent} above the minimum of the dip, and $t_2$ is the point in time when the intensity has decreased by \SI{75}{\percent} from the maximum of the overshoot. An illustration is given in Fig.~\ref{fig_hole_duration_illustration}. 
The magnitudes of the overshoots as well as the dips are measured from the local mean of the intensity trace obtained by low-pass filtering the intensity trace at \SI{100}{\mega\hertz}.
\\
The algorithm detects all local minima which deviate from the local mean by at least \SI{50}{\percent} and have a minimum distance of \SI{40}{\pico\second} to neighboring peaks. Starting from each detected dip, it identifies the point $t_1$ by moving backwards in time until the intensity has increased by \SI{75}{\percent} from the minimum of the dip. The point $t_2$ is found by moving forward in time: First, the algorithm moves from the dip to the intersection with the local mean and the intensity trace. Starting from this point the overshoot is tracked until the next intersection with the local mean. Based on this curve, the first local maximum is determined, and from there the algorithm moves to $t_2$ when the intensity has decreased by \SI{75}{\percent} from the local maximum. The algorithm requires a smooth intensity trace since for optimal performance the analyzed curves should be de- or increasing monotonously in the time window of observation. Therefore, the simulated intensity trace was additionally smoothed with a Gaussian filter with a FWHM of \SI{200}{\giga\hertz} in order to remove the impact of ASE noise. The measured intensity traces were up-sampled by a factor of ten to remove a discrete timing jitter in the hole duration caused by the finite sampling rate of the real-time oscilloscope of $\SI{6.25}{\pico\second}$ which is also the width of the bins in the histograms of the hole duration, which are presented in the subsequent section.
\\
\textbf{Results:}
\begin{figure}[t!]
	\centering
	\resizebox{0.48\textwidth}{!}{
		\includegraphics{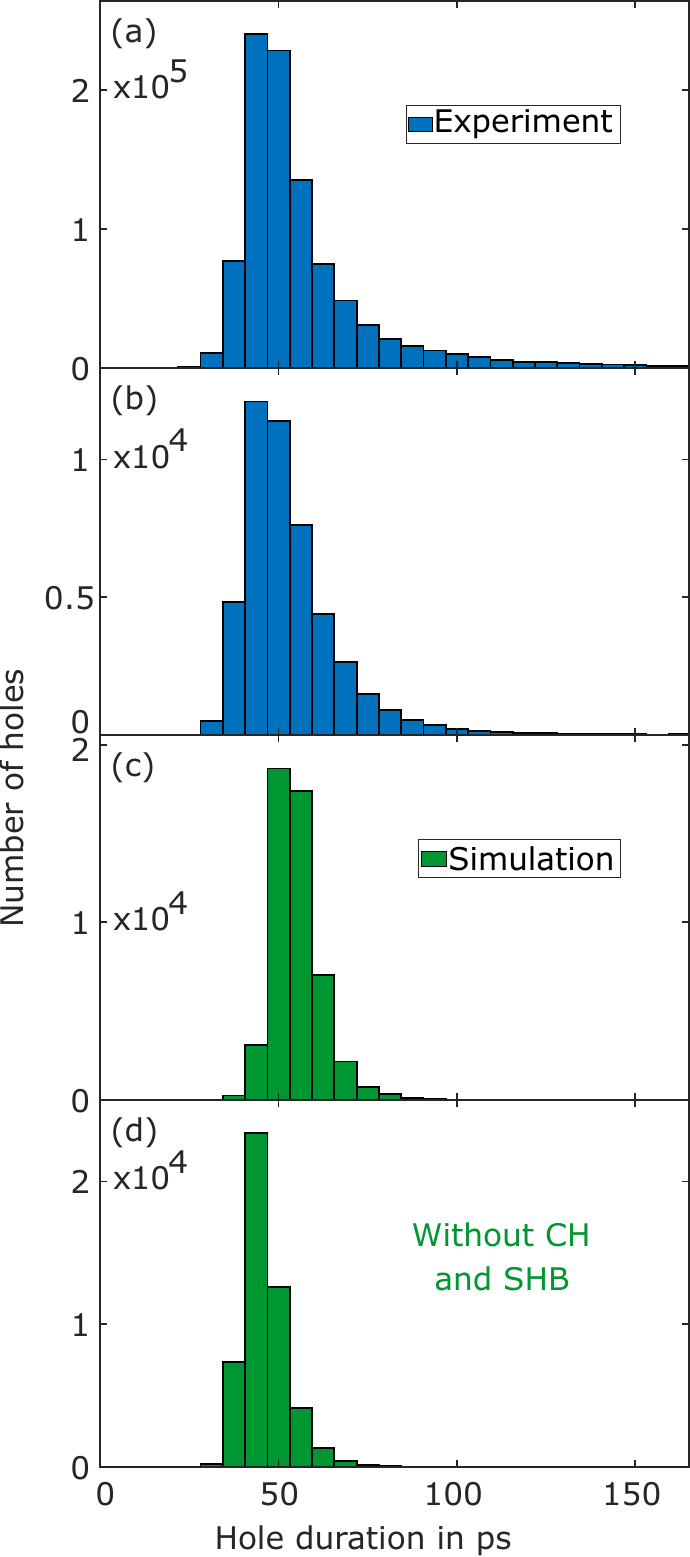}
	}
	\caption{(a) and (b) Histograms of the hole duration in experiment of backward sweeps detuned by (a) \SI{-5}{\hertz} and (b) \SI{-200}{\milli\hertz}, obtained from \SI{800}{} consecutive roundtrips. (c) and (d) Histograms computed from simulated intensity traces of the same setup as in (b) with the effects of CH and SHB included in (c) and excluded in (d).}
	\label{fig_histogram_holes}
\end{figure}
We evaluated \SI{800}{} consecutive backward sweeps of the second setup, corresponding to an observation time of \SI{2}{\milli\second}. The laser was detuned by \SI{-200}{\milli\hertz} as well as \SI{-5}{\hertz} to operate in the hole dominated regime. The center wavelength was \SI{1300}{\nano\meter} and the sweep bandwidth was \SI{100}{\nano\meter}. The histogram obtained from the experimental data of a laser detuned by \SI{-5}{\hertz} with the above discussed algorithm is shown in Fig.~\ref{fig_histogram_holes} (a). 
It can be observed that the hole duration is not a constant quantity and has an asymmetric distribution around a dominating peak which is in Fig.~\ref{fig_histogram_holes} (a) associated with the bar in the interval [\SI{40.625}{\pico\second} \SI{46.875}{\pico\second}], containing \SI{240406}{} holes which corresponds to \SI{25.24}{\percent}. A histogram produced from the same laser but detuned by \SI{-200}{\milli\hertz} has almost identical relative heights as demonstrated in Fig.~\ref{fig_histogram_holes} (b). Hence, it can be expected that the strength of the detuning is not a crucial parameter as long as the intensity trace is hole dominated. In addition, the distribution can be expected to be stationary. The maximum peak is also in the interval [\SI{40.625}{\pico\second} \SI{46.875}{\pico\second}] with \SI{12 103}{} holes and the estimated probability of \SI{25.21}{\percent} is similar as before. The reproduced histogram obtained from the simulation of the same setup as in Fig.~\ref{fig_histogram_holes} (b) is presented in Fig.~\ref{fig_histogram_holes} (c). 
The location of the peak, which is in Fig.~\ref{fig_histogram_holes} (c) in the interval [\SI{46.875}{\pico\second} \SI{53.125}{\pico\second}] with \SI{21860}{} holes (\SI{37.34}{\percent}), and the asymmetric broadening can be well reproduced. The variation in the hole duration can therefore be attributed to the inherent laser dynamics, particularly the frequency-shift-gain-loss interplay.\\ 
The more pronounced spread around the peak interval in the experimental data can partly be attributed to the wavelength and carrier density dependency of the carrier lifetime $\tau_\mathrm{c}$ or the linewidth enhancement factor $\alpha_\mathrm{N}$, which is not included in the simulation and thus the time dependency of $\tau_\mathrm{c}$ and $\alpha_\mathrm{N}$ as mentioned in Section~\ref{sec_CompToSimModel}. Furthermore, the bandwidth of the tunable bandpass filter is wavelength dependent and kept constant in the simulation. The bandwidth has been measured in the second setup and it is found to vary by \SI{20}{\percent} over most of the wavelength range. Our simulations show that the position of the peak of the histogram depends on the filter bandwidth and likely results in a broadening of the histogram when varying over the sweep. The above mentioned effects are expected to contribute to the additional broadening seen in experiment from Figs.~\ref{fig_histogram_holes} (a) and (b), yet require complex models prohibiting practical simulation times.\\
The percentage of detected holes which are larger than \SI{165}{\pico\second} and are thus not shown is \SI{<1}{\percent} in the experimental as well as the simulated data sets from Fig.~\ref{fig_histogram_holes}. The total number of holes is in Fig.~\ref{fig_histogram_holes} (a) \SI{952330}{}, in Fig.~\ref{fig_histogram_holes} (b) \SI{48037}{} and in the simulation in Figs.~\ref{fig_histogram_holes}  (c) and (d) \SI{50000}{} were collected.  
\\
The influence of the gain compression parameters on the hole duration statistics is displayed in Fig.~\ref{fig_histogram_holes} (d). Here, the effects of CH and SHB are not considered in the simulation, i.e., $\epsilon_{\mathrm{CH,SHB}} = \alpha_\mathrm{CH} = 0$. It can be seen that the asymmetry as well as the overall shape match better to experiment if the CH and SHB terms are included in the simulation. Based on these findings, it can be concluded that the nonlinearity added by CH and SHB contributes to a better agreement in the asymmetric broadening discussed above. All in all, a comparison to experimental results heavily relies on an accurate modeling of the frequency-shift-gain-loss interplay, and the hole duration statistics can serve as a tool to assess complex processes.
\\
The histogram of the hole duration presented above gives a valuable insight in the dynamical quantities of FDML lasers and can also be used to quantify simulation models and their accuracy. The key in this context is that the sign and the sweep direction determine the shape of the intensity patterns and one can control if the intensity trace is dominated by hole formation up to an upper bound when the detuning becomes too large, as shown in Fig.~\ref{fig_10Hz}. Our results are at least valid for the presented sweep parameters in this work. In other setups with an extremely narrow bandwidth of the bandpass filter of \SI{8.75}{\giga\hertz} it has been shown that also modified hole-type patterns can occur near the lasing threshold~\cite{gowda2020turbulent}.\\
The distribution of the hole duration is also of fundamental interest since the SOA gain medium with its complex microscopic dynamics has a significant impact on the shape of the intensity patterns, as discussed above. Therefore, a qualitative agreement of experimental and simulation data is a necessary condition for the simulation model to be accurate. 

\section{Conclusion}\label{sec_Conclusion}
We presented various intensity pattern types in rapidly swept FDML lasers and identified their operating conditions. By detuning the laser from the sweep rate at which the laser operates in the ultra-stable regime, the sign as well as the magnitude of the detuning and the sweep direction determine if the intensity trace of the laser is dominated by holes, localized fringes or short-term quasi-periodic groupings of pulsed patterns. Our experimental results are consistent with numerical simulations, showing that the formation of patterns is a result of the frequency-shift-gain-loss interplay in the laser cavity over a long time scale. By controlling the shape of the patterns, we were able to extract the distribution of the hole duration, giving insight in the dynamical quantities of the complex dynamics in FDML lasers.


\section*{Funding}
This work was supported by the German Research Foundation (DFG)(JI 115/4-2, JI 115/8-1, HU 1006/6 and EXC 2167--390884018), the European Union (ERC)(CoG no. 646669), the German Federal Ministry of Education and Research (BMBF)(no. 13GW0227B: ``Neuro-OCT'') and the State of Schleswig-Holstein, Germany (Excellence Chair Program by the universities of Kiel and L\"ubeck).
\section*{Disclosures}
Tom Pfeiffer: Optores GmbH (Employment, Patent, Recipient), Robert Huber: Optores GmbH (Personal Financial Interest, Patent, Recipient), Optovue Inc. (Patent, Recipient), Zeiss Meditec (Patent, Recipient), Abott (Patent, Recipient).
\section*{Data availability statement}
The datasets generated and analysed during the current study are available from the corresponding author on reasonable request.
%
%
%
%

\bibliographystyle{IEEEtran} 
\bibliography{references_APB}

\begin{thebibliography}{10}
\providecommand{\url}[1]{#1}
\csname url@samestyle\endcsname
\providecommand{\newblock}{\relax}
\providecommand{\bibinfo}[2]{#2}
\providecommand{\BIBentrySTDinterwordspacing}{\spaceskip=0pt\relax}
\providecommand{\BIBentryALTinterwordstretchfactor}{4}
\providecommand{\BIBentryALTinterwordspacing}{\spaceskip=\fontdimen2\font plus
\BIBentryALTinterwordstretchfactor\fontdimen3\font minus
  \fontdimen4\font\relax}
\providecommand{\BIBforeignlanguage}[2]{{%
\expandafter\ifx\csname l@#1\endcsname\relax
\typeout{** WARNING: IEEEtran.bst: No hyphenation pattern has been}%
\typeout{** loaded for the language `#1'. Using the pattern for}%
\typeout{** the default language instead.}%
\else
\language=\csname l@#1\endcsname
\fi
#2}}
\providecommand{\BIBdecl}{\relax}
\BIBdecl

\bibitem{huber2006buffered}
R.~Huber, D.~C. Adler, and J.~G. Fujimoto, ``Buffered {F}ourier domain mode
  locking: unidirectional swept laser sources for optical coherence tomography
  imaging at 370,000 lines/s,'' \emph{Optics Letters}, vol.~31, no.~20, pp.
  2975--2977, 2006.

\bibitem{wieser20105MHzOCT}
W.~Wieser, B.~R. Biedermann, T.~Klein, C.~M. Eigenwillig, and R.~Huber,
  ``Multi-megahertz {OCT}: High quality 3{D} imaging at 20 million {A}-scans
  and 4.5 {GV}oxels per second,'' \emph{Optics Express}, vol.~18, no.~14, pp.
  14\,685--14\,704, 2010.

\bibitem{klein2011megahertz}
T.~Klein, W.~Wieser, C.~M. Eigenwillig, B.~R. Biedermann, and R.~Huber,
  ``Megahertz {OCT} for ultrawide-field retinal imaging with a 1050nm {F}ourier
  domain mode-locked laser,'' \emph{Optics Express}, vol.~19, no.~4, pp.
  3044--3062, 2011.

\bibitem{klein2013multi}
T.~Klein, W.~Wieser, L.~Reznicek, A.~Neubauer, A.~Kampik, and R.~Huber,
  ``Multi-{MH}z retinal {OCT},'' \emph{Biomedical Optics Express}, vol.~4,
  no.~10, pp. 1890--1908, 2013.

\bibitem{kolb2016megahertz}
J.~P. Kolb, T.~Klein, M.~Eibl, T.~Pfeiffer, W.~Wieser, and R.~Huber,
  ``Megahertz {FDML} laser with up to 143nm sweep range for ultrahigh
  resolution {OCT} at 1050nm,'' in \emph{Optical Coherence Tomography and
  Coherence Domain Optical Methods in Biomedicine XX}, vol. 9697.\hskip 1em
  plus 0.5em minus 0.4em\relax International Society for Optics and Photonics,
  2016, p. 969703.

\bibitem{kolb2018high}
J.~P. Kolb, T.~Pfeiffer, M.~Eibl, H.~Hakert, and R.~Huber, ``High-resolution
  retinal swept source optical coherence tomography with an ultra-wideband
  {F}ourier-domain mode-locked laser at {MHz A-scan} rates,'' \emph{Biomedical
  Optics Express}, vol.~9, no.~1, pp. 120--130, 2018.

\bibitem{biedermann2009recent}
B.~R. Biedermann, W.~Wieser, C.~M. Eigenwillig, and R.~Huber, ``Recent
  developments in {F}ourier {D}omain {M}ode {L}ocked lasers for optical
  coherence tomography: {I}maging at 1310 nm vs. 1550 nm wavelength,''
  \emph{Journal of Biophotonics}, vol.~2, no. 6-7, pp. 357--363, 2009.

\bibitem{reznicek2014megahertz}
L.~Reznicek, T.~Klein, W.~Wieser, M.~Kernt, A.~Wolf, C.~Haritoglou, A.~Kampik,
  R.~Huber, and A.~S. Neubauer, ``Megahertz ultra-wide-field swept-source
  retina optical coherence tomography compared to current existing imaging
  devices,'' \emph{Graefe's Archive for Clinical and Experimental
  Ophthalmology}, vol. 252, no.~6, pp. 1009--1016, 2014.

\bibitem{hartl2001ultrahigh}
I.~Hartl, X.~D. Li, C.~Chudoba, R.~K. Ghanta, T.~H. Ko, J.~G. Fujimoto, J.~K.
  Ranka, and R.~S. Windeler, ``Ultrahigh-resolution optical coherence
  tomography using continuum generation in an air--silica microstructure
  optical fiber,'' \emph{Optics Letters}, vol.~26, no.~9, pp. 608--610, 2001.

\bibitem{bourquin2003ultrahigh}
S.~Bourquin, A.~D. Aguirre, I.~Hartl, P.~Hsiung, T.~H. Ko, J.~G. Fujimoto,
  T.~A. Birks, W.~J. Wadsworth, U.~B{\"u}nting, and D.~Kopf, ``Ultrahigh
  resolution real time {OCT} imaging using a compact femtosecond {Nd:Glass}
  laser and nonlinear fiber,'' \emph{Optics Express}, vol.~11, no.~24, pp.
  3290--3297, 2003.

\bibitem{nishizawa2004real}
N.~Nishizawa, Y.~Chen, P.~Hsiung, E.~P. Ippen, and J.~G. Fujimoto, ``Real-time,
  ultrahigh-resolution, optical coherence tomography with an all-fiber,
  femtosecond fiber laser continuum at 1.5 $\mu$m,'' \emph{Optics Letters},
  vol.~29, no.~24, pp. 2846--2848, 2004.

\bibitem{leitgeb2004ultrahigh}
R.~A. Leitgeb, W.~Drexler, A.~Unterhuber, B.~Hermann, T.~Bajraszewski, T.~Le,
  A.~Stingl, and A.~F. Fercher, ``Ultrahigh resolution {F}ourier domain optical
  coherence tomography,'' \emph{Optics Express}, vol.~12, no.~10, pp.
  2156--2165, 2004.

\bibitem{wojtkowski2004ultrahigh}
M.~Wojtkowski, V.~J. Srinivasan, T.~H. Ko, J.~G. Fujimoto, A.~Kowalczyk, and
  J.~S. Duker, ``Ultrahigh-resolution, high-speed, {F}ourier domain optical
  coherence tomography and methods for dispersion compensation,'' \emph{Optics
  Express}, vol.~12, no.~11, pp. 2404--2422, 2004.

\bibitem{potsaid2010ultrahigh}
B.~Potsaid, B.~Baumann, D.~Huang, S.~Barry, A.~E. Cable, J.~S. Schuman, J.~S.
  Duker, and J.~G. Fujimoto, ``Ultrahigh speed 1050nm swept source / {F}ourier
  domain {OCT} retinal and anterior segment imaging at 100,000 to 400,000 axial
  scans per second,'' \emph{Optics Express}, vol.~18, no.~19, pp.
  20\,029--20\,048, 2010.

\bibitem{nishizawa2018wavelength}
N.~Nishizawa, H.~Kawagoe, M.~Yamanaka, M.~Matsushima, K.~Mori, and T.~Kawabe,
  ``Wavelength dependence of ultrahigh-resolution optical coherence tomography
  using supercontinuum for biomedical imaging,'' \emph{IEEE Journal of Selected
  Topics in Quantum Electronics}, vol.~25, no.~1, pp. 1--15, 2018.

\bibitem{wartak2020micro}
A.~Wartak, M.~S. Schenk, V.~B{\"u}hler, S.~A. Kassumeh, R.~Birngruber, and
  G.~J. Tearney, ``Micro-optical coherence tomography for high-resolution
  morphologic imaging of cellular and nerval corneal micro-structures,''
  \emph{Biomedical Optics Express}, vol.~11, no.~10, pp. 5920--5933, 2020.

\bibitem{butler2015single}
T.~Butler, S.~Slepneva, B.~O’Shaughnessy, B.~Kelleher, D.~Goulding,
  S.~Hegarty, H.-C. Lyu, K.~Karnowski, M.~Wojtkowski, and G.~Huyet, ``Single
  shot, time-resolved measurement of the coherence properties of {OCT} swept
  source lasers,'' \emph{Optics Letters}, vol.~40, no.~10, pp. 2277--2280,
  2015.

\bibitem{biedermann2010direct}
B.~R. Biedermann, W.~Wieser, C.~M. Eigenwillig, T.~Klein, and R.~Huber,
  ``Direct measurement of the instantaneous linewidth of rapidly
  wavelength-swept lasers,'' \emph{Optics Letters}, vol.~35, no.~22, pp.
  3733--3735, 2010.

\bibitem{todor2011instantaneous}
S.~Todor, B.~Biedermann, W.~Wieser, R.~Huber, and C.~Jirauschek,
  ``Instantaneous lineshape analysis of {F}ourier domain mode-locked lasers,''
  \emph{Optics Express}, vol.~19, no.~9, pp. 8802--8807, 2011.

\bibitem{kraetschmer2009ultrastable}
T.~Kraetschmer and S.~T. Sanders, ``Ultrastable {F}ourier domain mode locking
  observed in a laser sweeping 1363.8 -- 1367.3 nm,'' in \emph{Conference on
  Lasers and Electro-Optics}.\hskip 1em plus 0.5em minus 0.4em\relax Optical
  Society of America, 2009, p. CFB4.

\bibitem{slepneva2013dynamics}
S.~Slepneva, B.~Kelleher, B.~O'Shaughnessy, S.~P. Hegarty, A.~G. Vladimirov,
  and G.~Huyet, ``Dynamics of {F}ourier domain mode-locked lasers,''
  \emph{Optics Express}, vol.~21, no.~16, pp. 19\,240--19\,251, 2013.

\bibitem{slepneva2019convective}
S.~Slepneva, B.~O'Shaughnessy, A.~G. Vladimirov, S.~Rica, E.~A. Viktorov, and
  G.~Huyet, ``Convective {N}ozaki-{B}ekki holes in a long cavity {OCT} laser,''
  \emph{Optics Express}, vol.~27, no.~11, pp. 16\,395--16\,404, 2019.

\bibitem{pfeiffer2018ultra}
T.~Pfeiffer, M.~Petermann, W.~Draxinger, C.~Jirauschek, and R.~Huber, ``Ultra
  low noise {F}ourier domain mode locked laser for high quality megahertz
  optical coherence tomography,'' \emph{Biomedical Optics Express}, vol.~9,
  no.~9, pp. 4130--4148, 2018.

\bibitem{lippok2019extended}
N.~Lippok, M.~Siddiqui, B.~J. Vakoc, and B.~E. Bouma, ``Extended coherence
  length and depth ranging using a {F}ourier-domain mode-locked frequency comb
  and circular interferometric ranging,'' \emph{Physical Review Applied},
  vol.~11, no.~1, p. 014018, 2019.

\bibitem{jung2008characterization}
E.~J. Jung, C.-S. Kim, M.~Y. Jeong, M.~K. Kim, M.~Y. Jeon, W.~Jung, and
  Z.~Chen, ``Characterization of {F}{B}{G} sensor interrogation based on a
  {F}{D}{M}{L} wavelength swept laser,'' \emph{Optics Express}, vol.~16,
  no.~21, pp. 16\,552--16\,560, 2008.

\bibitem{marschall2010fourier}
S.~Marschall, T.~Klein, W.~Wieser, B.~R. Biedermann, K.~Hsu, K.~P. Hansen,
  B.~Sumpf, K.-H. Hasler, G.~Erbert, O.~B. Jensen, C.~Pedersen, R.~Huber, and
  P.~E. Andersen, ``{F}ourier domain mode-locked swept source at 1050 nm based
  on a tapered amplifier,'' \emph{Optics Express}, vol.~18, no.~15, pp.
  15\,820--15\,831, 2010.

\bibitem{schmidt2020self}
M.~Schmidt, T.~Pfeiffer, C.~Grill, R.~Huber, and C.~Jirauschek,
  ``Self-stabilization mechanism in ultra-stable {F}ourier domain mode-locked
  ({FDML}) lasers,'' \emph{OSA Continuum}, vol.~3, no.~6, pp. 1589--1607, 2020.

\bibitem{pimenov2017dispersive}
A.~Pimenov, S.~Slepneva, G.~Huyet, and A.~G. Vladimirov, ``Dispersive
  time-delay dynamical systems,'' \emph{Physical Review Letters}, vol. 118,
  no.~19, p. 193901, 2017.

\bibitem{li2017eckhaus}
F.~Li, K.~Nakkeeran, J.~N. Kutz, J.~Yuan, Z.~Kang, X.~Zhang, and P.~K.~A. Wai,
  ``Eckhaus instability in the {F}ourier-domain mode locked fiber laser
  cavity,'' \emph{arXiv preprint arXiv:1707.08304}, 2017.

\bibitem{jirauschek2009theoretical}
C.~Jirauschek, B.~Biedermann, and R.~Huber, ``A theoretical description of
  {F}ourier domain mode locked lasers,'' \emph{Optics Express}, vol.~17,
  no.~26, pp. 24\,013--24\,019, 2009.

\bibitem{wan2017time}
M.~Wan, F.~Li, X.~Feng, X.~Wang, Y.~Cao, B.-O. Guan, D.~Huang, J.~Yuan, and
  P.~K.~A. Wai, ``Time and {F}ourier domain jointly mode locked frequency comb
  swept fiber laser,'' \emph{Optics Express}, vol.~25, no.~26, pp.
  32\,705--32\,712, 2017.

\bibitem{mecozzi1997saturationFWM}
A.~Mecozzi and J.~M{\o}rk, ``Saturation effects in nondegenerate four-wave
  mixing between short optical pulses in semiconductor laser amplifiers,''
  \emph{IEEE Journal of Selected Topics in Quantum Electronics}, vol.~3, no.~5,
  pp. 1190--1207, 1997.

\bibitem{wang2007temporal}
J.~Wang, A.~Maitra, C.~G. Poulton, W.~Freude, and J.~Leuthold, ``Temporal
  dynamics of the alpha factor in semiconductor optical amplifiers,''
  \emph{Journal of Lightwave Technology}, vol.~25, no.~3, pp. 891--900, 2007.

\bibitem{zilkie2008time}
A.~J. Zilkie, J.~Meier, M.~Mojahedi, A.~S. Helmy, P.~J. Poole, P.~Barrios,
  D.~Poitras, T.~J. Rotter, C.~Yang, A.~Stintz, K.~J. Malloy, P.~W.~E. Smith,
  and J.~S. Aitchison, ``Time-resolved linewidth enhancement factors in quantum
  dot and higher-dimensional semiconductor amplifiers operating at 1.55
  $\mu$m,'' \emph{Journal of Lightwave Technology}, vol.~26, no.~11, pp.
  1498--1509, 2008.

\bibitem{cassioli2000time}
D.~Cassioli, S.~Scotti, and A.~Mecozzi, ``A time-domain computer simulator of
  the nonlinear response of semiconductor optical amplifiers,'' \emph{IEEE
  Journal of Quantum Electronics}, vol.~36, no.~9, pp. 1072--1080, 2000.

\bibitem{corless1996lambertw}
R.~M. Corless, G.~H. Gonnet, D.~E.~G. Hare, D.~J. Jeffrey, and D.~E. Knuth,
  ``On the {L}ambert{W} function,'' \emph{Advances in Computational
  Mathematics}, vol.~5, no.~1, pp. 329--359, 1996.

\bibitem{grelu2012dissipative}
P.~Grelu and N.~Akhmediev, ``Dissipative solitons for mode-locked lasers,''
  \emph{Nature Photonics}, vol.~6, no.~2, pp. 84--92, 2012.

\bibitem{liao2005clustering}
T.~W. Liao, ``Clustering of time series data—a survey,'' \emph{Pattern
  Recognition}, vol.~38, no.~11, pp. 1857--1874, 2005.

\bibitem{todor2012balance}
S.~Todor, B.~Biedermann, R.~Huber, and C.~Jirauschek, ``Balance of physical
  effects causing stationary operation of {F}ourier domain mode-locked
  lasers,'' \emph{Journal of the Optical Society of America B}, vol.~29, no.~4,
  pp. 656--664, 2012.

\bibitem{girard2011soa}
S.~L. Girard, M.~Pich{\'e}, H.~Chen, G.~W. Schinn, W.-Y. Oh, and B.~E. Bouma,
  ``{SOA} fiber ring lasers: {S}ingle-versus multiple-mode oscillation,''
  \emph{IEEE Journal of Selected Topics in Quantum Electronics}, vol.~17,
  no.~6, pp. 1513--1520, 2011.

\bibitem{gowda2020turbulent}
U.~Gowda, A.~Roche, A.~Pimenov, A.~G. Vladimirov, S.~Slepneva, E.~A. Viktorov,
  and G.~Huyet, ``Turbulent coherent structures in a long cavity semiconductor
  laser near the lasing threshold,'' \emph{Optics Letters}, vol.~45, no.~17,
  pp. 4903--4906, 2020.

\end{thebibliography}

\end{document}